\documentclass[aps,prl,twocolumn,groupedaddress]{revtex4-1}
\usepackage{graphicx}
\usepackage{amsmath}
\usepackage{amssymb}
\usepackage{float}
\usepackage{bm,times}
\usepackage{graphicx}
\usepackage{pdfpages}

\begin{document}
\newcommand{\bra}[1]{\mbox{\ensuremath{\langle #1 \vert}}}
\newcommand{\ket}[1]{\mbox{\ensuremath{\vert #1 \rangle}}}
\newcommand{\mb}[1]{\mathbf{#1}}
\newcommand{\phipp}{\big|\phi_{\mb{p}}^{(+)}\big>}
\newcommand{\phipav}{\big|\phi_{\mb{p}}^{\p{av}}\big>}
\newcommand{\pp}[1]{\big|\psi_{p}(#1)\big>}
\newcommand{\drdy}[1]{\sqrt{-R'(#1)}}
\newcommand{\Rb}{$^{87}$Rb}
\newcommand{\kf}{$^{40}$K}
\newcommand{\na}{${^{23}}$Na}
\newcommand{\muK}{\:\mu\textrm{K}}
\newcommand{\p}[1]{\textrm{#1}}
\newcommand\T{\rule{0pt}{2.6ex}}
\newcommand\B{\rule[-1.2ex]{0pt}{0pt}}
\newcommand{\reffig}[1]{\mbox{Fig.~\ref{#1}}}
\newcommand{\refeq}[1]{\mbox{Eq.~(\ref{#1})}}
\newcommand{\nuke}[1]{}
\newcommand{\note}[1]{\textcolor{red}{[\textrm{#1}]}}

\title{Cooling a single atom in an optical tweezer to its quantum ground state}

\author{A. M. Kaufman}
\author{B. J. Lester}
\author{C. A. Regal}
\email[E-mail: ]{regal@jila.colorado.edu}
\address{JILA, University of Colorado and National Institute of Standards and Technology, and
Department of Physics, University of Colorado, Boulder, Colorado 80309, USA}

\begin{abstract} 
We report cooling of a single neutral atom to its three-dimensional vibrational ground state in an optical tweezer. After employing Raman sideband cooling for tens of milliseconds, we measure via sideband spectroscopy a three-dimensional ground-state occupation of $\sim\!\!90\%$. We further observe coherent control of the spin and motional state of the trapped atom. Our demonstration shows that an optical tweezer, formed simply by a tightly focused beam of light, creates sufficient confinement for efficient sideband cooling. This source of ground-state neutral atoms will be instrumental in numerous quantum simulation and logic applications that require a versatile platform for storing and manipulating ultracold single neutral atoms. For example, these results will improve current optical tweezer experiments studying atom-photon coupling and Rydberg quantum logic gates, and could provide new opportunities such as rapid production of single dipolar molecules or quantum simulation in tweezer arrays.
\end{abstract}

\date{\today}

\maketitle

Trapped ultracold neutral atoms provide a promising starting point for quantum simulation and computation.  Ideally, experiments would be able to initialize a homogeneous array of atoms in an arbitrary quantum state, reconfigure this array in real time, and turn on and off interactions. Optical tweezers represent an interesting route towards this vision because they have the capability to integrate multi-qubit storage, read-out, and transport~\cite{Schlosser2001, Beugnon2007, Zhang2010,Wilk2010}.  Recently, an idea for near-deterministic loading of optical tweezer traps has been demonstrated~\cite{Grunzweig2010}.  By combining this idea with the capability to detect and correct occupation defects, it is a near-term possibility to build an ordered neutral-atom array atom by atom~\cite{Weiss2004}.  However, an unrealized prerequisite for this vision of a low-entropy tweezer array is the ability to place a specific single atom in its vibrational ground state.  Such control of single-particle motion has been an integral part of trapped-ion experiments for decades.  In this article, we show that analogous control can be realized with a neutral atom.  We optically cool a single \Rb~atom in a tweezer to its three-dimensional vibrational ground state with $90\%$ probability, which is a two order of magnitude improvement upon previous experiments employing laser-cooled atoms in optical tweezers.

\begin{figure}[ht]
	\centering
	\includegraphics[scale=0.43]{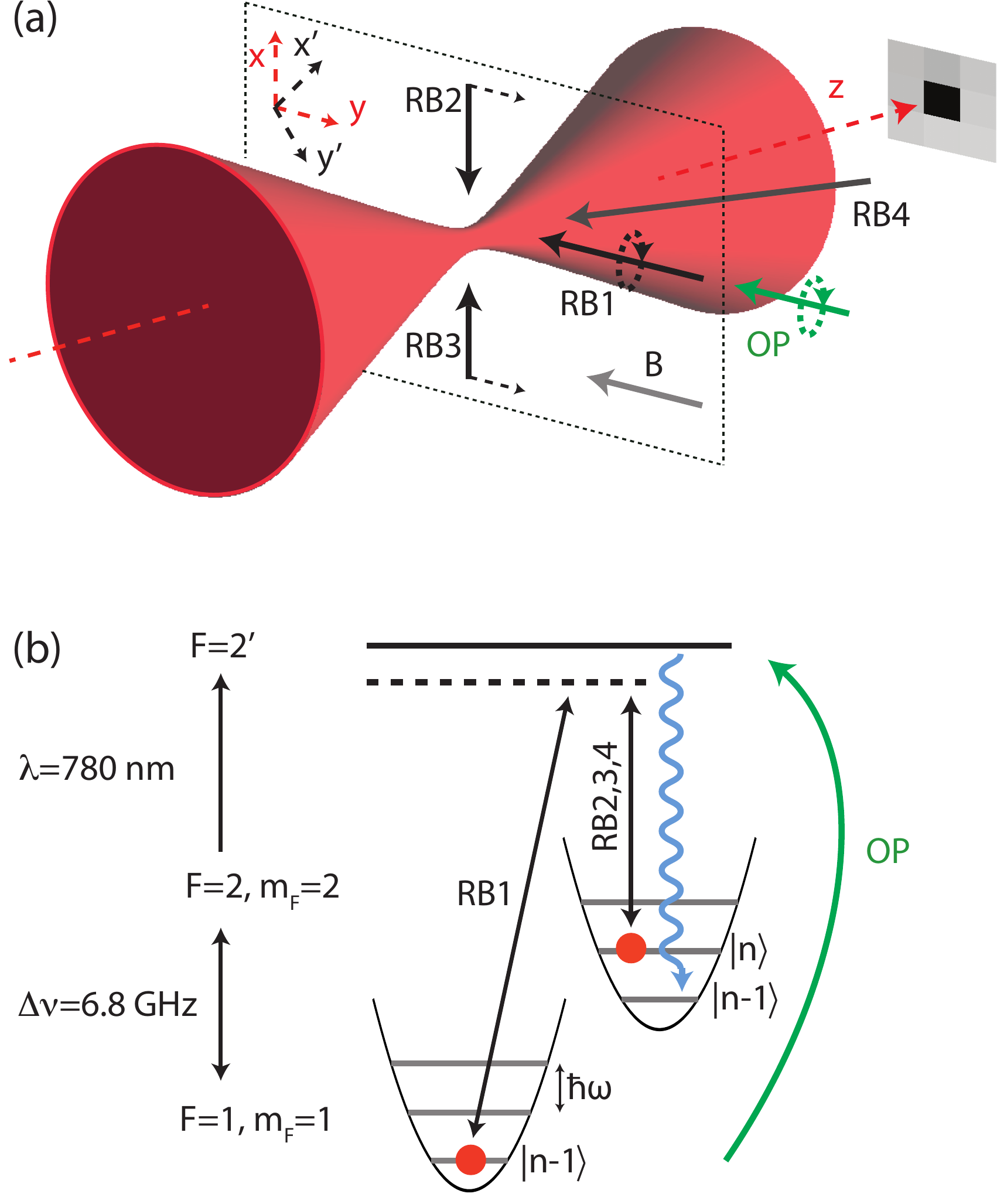}
	\caption{Experimental setup for tweezer trap, detection, and three-dimensional motional control. (a) An optical tweezer created with a high NA objective lens traps a single neutral atom, and the atom is imaged along the $z$-axis with the same lens.  Orthogonal radial axes, indicated by $x'$ and $y'$, are addressed by Raman beam 1 (RB1) ($\sigma^+$-polarized) and RB2 ($\pi$-polarized), or RB1 and RB3  ($\pi$-polarized).  RB1 and RB4 (linearly polarized in y-z plane) address the axial direction. Note we should be able to cool all three axes with a single pair of counterpropagating beams.  (b) Level diagram for \Rb~with associated beams from (a).  The Raman light is $50$ GHz red detuned of the excited state manifold.  Optical pumping (OP) consists of repump light on the $F=1 \rightarrow 2'$ transition along with optical pumping light on $F=2 \rightarrow 2'$.}
	\label{fig:exp}
\end{figure}

Motional control of neutral atoms has a rich history, and increasingly interest has turned to the problem of single-atom control. To date, optical lattices created by standing waves of light have been the main platform for realizing motional control of collections of single neutral atoms. Approaches have included dramatic demonstrations of the Mott insulator transition of an evaporatively cooled gas~\cite{Bakr2009, Weitenberg2011}, and exploration of laser cooling collections of atoms in a lattice~\cite{Hamann1997,Perrin1998,Vuletic1998,Han2000,Maunz2004,Boozer2006,Forster2009,Blatt2009,Li2012}. Most recently, in spin-dependent lattices experimenters have harnessed microwave signals to cool and control atomic motion~\cite{Forster2009,Li2012}.  Our work in an optical tweezer breaks with typical lattice experiments, and instead more closely resembles the sideband cooling and spectroscopy techniques used with atomic ions~\cite{Monroe1995}. We hold a single atom in a far-detuned tweezer trap and apply near-resonant, pulsed cooling and spectroscopy light that couples two ground state hyperfine levels (Fig.~\ref{fig:exp}(a)). The complete separation of the trapping and cooling beams~\cite{Han2000,Boozer2006, Blatt2009} allows us to realize rapid cooling as well as low trap spontaneous emission rates and hence long qubit coherence times.

Already, optical tweezers have been used to realize Rydberg quantum-logic gates and a variety of protocols coupling atoms and photons.  In these experiments the thermal motion of the atoms has caused deleterious effects, such as dynamic light shifts, mitigated atom-photon coupling, and dephasing of high fidelity Rydberg gates~\cite{Darque2005, Tey2008, Zhang2010,Wilk2010}. Attaining full three-dimensional motional control~\cite{Jochim2011, Esslinger2011} would not only strengthen current tweezer applications, but also expand their use to experiments that are currently considered only in the context of evaporatively cooled gases. For example, one could combine two traps and realize significant wavefunction overlap for Feshbach molecule association, and hence creation and control of single dipolar molecules~\cite{Ni2008,Jochim2011}. Or with sufficient tunneling between traps and control of a spin degree of freedom, one could study strongly correlated physics proposed in lattices~\cite{Paredes2008}  with a platform amenable to studying non-equilibrium dynamics.

The versatility of our tweezer trapping and cooling platform also presents new opportunities.  In particular, a frontier in quantum interfaces is realizing strong-coupling of neutral atoms to nanophotonic circuits~\cite{Aoki2006,Vetsch2010, Chang2009}, but trapping and controlling atoms in nanoscale, near-surface potentials remains a considerable challenge. Ground state cooling of an atom in an optical tweezer advances this frontier by both providing a mobile, highly localized reservoir of single neutral atoms for loading nanoscale potentials, and advancing techniques for in-situ cooling within these potentials.

We laser cool to the ground state by employing a technique known as Raman sideband cooling~\cite{KermanThesis,Leibfried2003}. Raman sideband cooling operates by iterating on a two step process that manipulates the atomic spin and motion, and resolves motional transitions even when the trap frequency is less than the electronic transition linewidth.  In our scheme, we utilize two hyperfine levels of $^{87}$Rb:  $\ket{F,m_F} \equiv \ket{2,2}$ and $\ket{1,1}$ (Fig.~\ref{fig:exp}(b)), where $F$ is the hyperfine angular momentum quantum number and $m_F$ its projection. The first step is a stimulated Raman transition on $\ket{2,2;n}\leftrightarrow\ket{1,1;n-1}$ that reduces the vibrational state by one quantum while also realizing a spin flip. The second step is the dissipative step: The atom is optically pumped back to the initial spin-state via a spontaneous Raman process, while the photon carries away entropy.  By repeating these steps, population accumulates in the $\ket{2,2;0}$ state because it is dark to the Raman beams and the optical pumping light. 

For Raman cooling to be successful, the optical pumping step must preserve the reduced vibrational state. The vibrational excitation probability depends on the motional quantum number $n$ and the Lamb-Dicke parameter, $\eta^{OP} \equiv x_0 k$, where $x_0 = (\hbar/2m \omega)^{1/2}$ is the oscillator length for a particle of mass $m$ and $k$ is the optical pumping wave number. Raman cooling begins with an atom in a mixed thermal state, with a temperature corresponding to an average vibrational quantum number $\bar{n}$.  In this case, the excitation probability due to a single scattered photon scales with $(\eta^{OP}_{\mathrm{eff}})^2 \equiv (2\bar{n} +1)(\eta^{OP})^2$, and hence the Raman cooling efficiency scales inversely with $(\eta^{OP}_{\mathrm{eff}})^2$.

Therefore, ground state cooling (achieving $\bar{n} = 0$) requires (1) low enough initial temperatures before starting Raman cooling, and (2) sufficient confinement, i.e. large trap frequencies.   To realize low initial temperatures in our optical tweezer trap, we carefully apply the sub-Doppler cooling technique of polarization gradient cooling (PGC)~\cite{Lett1988}.   To realize strong three-dimensional confinement we use a tightly focused optical tweezer trap; the trap is formed using far-off-resonant light at 852 nm and a 0.6 numerical aperture (NA) objective.  With $2.8~\mathrm{mW}$ of power in the central focal spot, we measure trap frequencies of $\{\omega_z, \omega_{x'}, \omega_{y'}\} /2\pi= \{30,154,150\}~\mathrm{kHz}$.  The large range of frequencies spanned by the axial ($z$) and radial ($x,y$) dimensions of our trap allow us to explore the challenges to robust Raman cooling that accompany increasing Lamb-Dicke parameters and initial occupations.  

Another challenge specific to the optical tweezer platform is effective magnetic fields (vector light shifts) induced by a linearly polarized dipole trap with a non-paraxial focus.  This additional field could (1) dephase field-sensitive transitions, (2) disrupt the quantization axis and hence optical pumping fidelity, and (3) increase realizable PGC temperatures.  For our NA, detuning, and typical intensity, we calculate that the effective magnetic field is $ \pm 0.13~\mathrm{G}$ over $\pm 50~\mathrm{nm}$, and points along the cross-product of the dipole trap axis ($z$) and its polarization ($y$)~\cite{Ewolf2}.  During Raman cooling, we set our quantization field to $3~\mathrm{G}$ in a direction orthogonal to the effective magnetic field, mitigating effects (1) and (2). 

Our experiment begins by loading the tweezer trap from a magneto-optical trap overlapped with the optical trap focus.  We use light-assisted collisions to realize zero or one atoms in the trap with approximately equal probability \cite{Schlosser2001}.  In order to post-select on the presence of one atom, we take an initial image of the atom immediately after the loading sequence.  We start the cooling sequence by applying PGC light for 5 ms in a $\sigma^+$-$\sigma^-$ configuration using three retro-reflected beams (see the Appendix). We address the challenging problem of stabilizing the phase of the interference pattern with respect to the sub-wavelength extent of the atom in the tweezer trap by modulating the position of our retro-reflecting mirrors at $1~\mathrm{kHz}$. This yields a time-averaged cooling that does not exhibit shot-to-shot variations in the final PGC temperature due to slow fluctuations in the interference pattern. 

\begin{figure*}[ht]
	\centering
	\includegraphics[scale = .33]{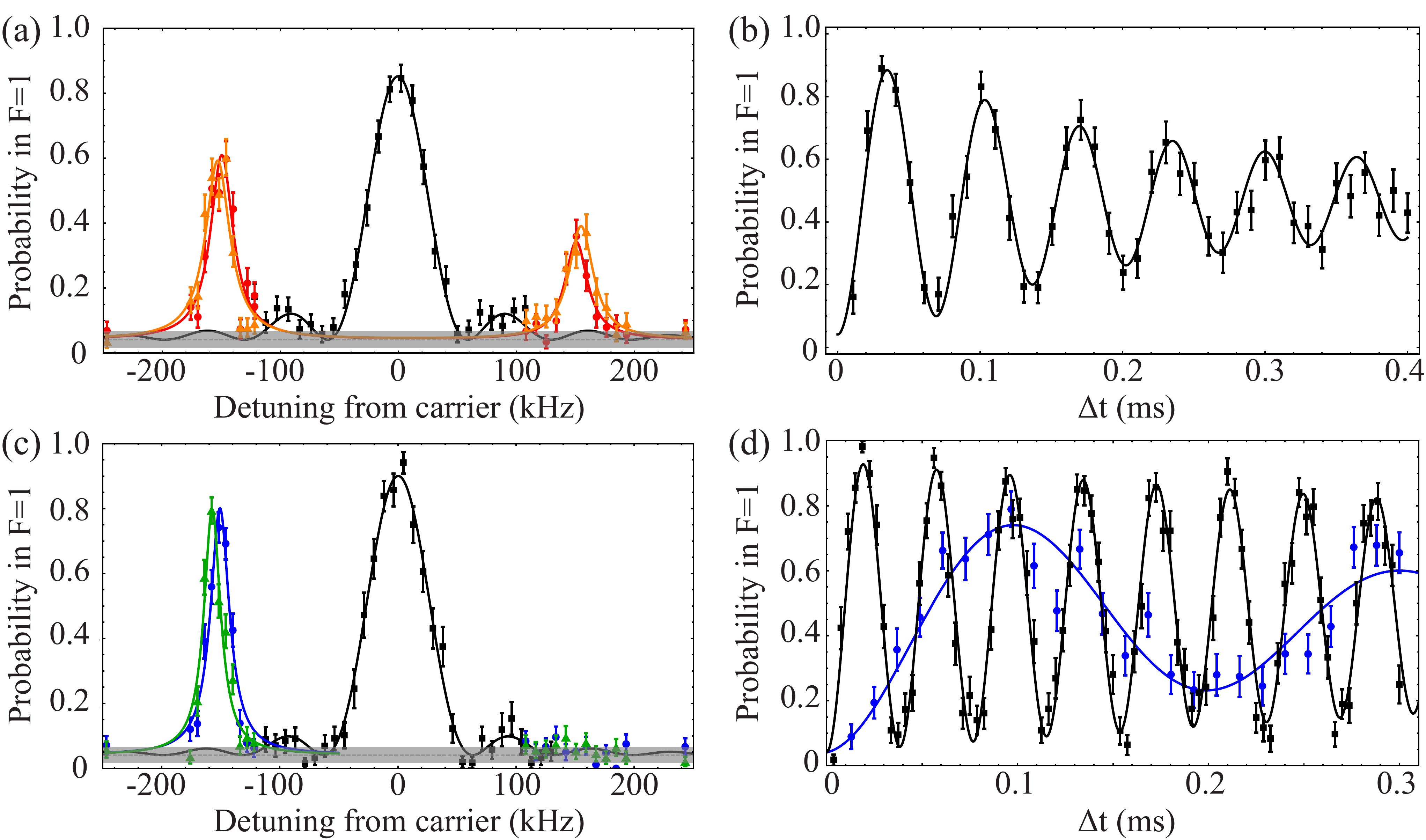}
	\caption{Single atom sideband spectra and Rabi oscillations in the radial dimensions before (a, b) and after (c, d) ground state cooling. (a) The black squares are a carrier peak in the $y'$ direction using a $\Delta t=15~\mathrm{\mu s}$ (near $\pi$) pulse.  The red circles (orange triangles) are sidebands along the $y'$ ($x'$) axis for a $75~\mathrm{\mu s}$ (near $\pi$) pulse, demonstrating an initial thermal population of vibrational states.  The solid lines are fits to a Rabi sinc function (See Supplementary Material for definition) for the carrier and Lorentzians (an approximation) for the sidebands; each fit contains an offset at our measured background (gray shaded region centered at 0.04).  (b) Carrier Rabi oscillations for the $y'$ direction showing dephasing of a thermal state.  Here the carrier Rabi frequency was set to 15 kHz, instead of the usual 26 kHz.  The solid line is a fit to the data using a thermal distribution of Rabi frequencies. (c) Raman cooled radial sidebands; no Raman cooling is applied to the axial direction for these data.  The black squares are a cooled carrier peak using a $15~\mathrm{\mu s}$ pulse.  The blue circles (green triangles) are spectra along the $y'$ ($x'$) axis using a $75~\mathrm{\mu s}$ pulse, displaying a significant asymmetry that is the hallmark of a large ground state population.  (d) Rabi oscillations for a radial ground state cooled atom with a fit to a damped sine for the carrier (black squares) and the $\Delta n =+1$ sideband (blue circles), which demonstrates coherent control of the spin-motional states; the carrier dephasing is suppressed due to the purity of the vibrational distribution. Each data point is an average of 150 experimental runs, and hence $\sim75$ atoms.}
	\label{fig:spectra}
\end{figure*} 

As our first application of Raman coupling, we diagnose a thermal occupation using sideband spectroscopy.  After PGC is performed, we optically pump to $\ket{F,m_F} \equiv \ket{2,2}$.  A single pair of Raman beams is chosen according to the dimension we wish to probe (Fig.~\ref{fig:exp}(a)), and we interrogate the atom with a square pulse of length $\Delta t$, leading to transitions to the $\ket{1,1}$ state when a resonance is satisfied.  The Lamb-Dicke parameter for the Raman transition, which quantifies the motional coupling, is given by $\eta^R\equiv x_0\Delta k$, where $\Delta k$ is the momentum transferred by the Raman process along the motional axis.  For our geometry $\eta^R=0.22$ for the radial dimension and $\eta^R=0.23$ for the axial dimension. 

Figure~\ref{fig:spectra}(a) shows sideband spectra after PGC along orthogonal radial directions. The asymmetry between the $\Delta n = -1$ and the $\Delta n = +1$ peaks determines the population of the atoms in the ground state. (Because we begin in the upper hyperfine state, the $\Delta n = -1$ peak is on the right in Fig.~\ref{fig:spectra}.)  By equating the ratio of the $\Delta n = -1$ and $\Delta n = +1$ sidebands to $\bar{n}/(\bar{n}+1)$ we find $\bar{n}_{y'} = 1.1 \pm 0.4$ or $\mathrm{T_{y'}} = 11 \pm 3~\mathrm{\mu K}$ assuming a thermal population distribution; correspondingly $\bar{n}_{x'} = 1.7 \pm 0.7$. Using the dephasing of the carrier Rabi oscillations in these data, we can also extract a temperature via the coherent evolution of the thermal state [Fig.~\ref{fig:spectra}(b)] to find $\mathrm{T_{y'}} \leq 16 \pm 2~\mu \mathrm{K}$~\cite{note1}.  For comparison, we also employ a standard thermometry technique in which the atom is quickly released, and the probability of recapturing the atom at a variable time later is measured and compared to a classical Monte Carlo model \cite{Tuchendler2008} from which we estimate a temperature of $13 \pm 1~\mu \mathrm{K}$.  The agreement between our three measurements validates sideband spectroscopy as a reliable form of thermometry in an optical tweezer, and we find low PGC temperatures are possible despite the varying effective magnetic fields within the focus.

\begin{figure*}[ht]
	\centering
	\includegraphics[scale=0.32]{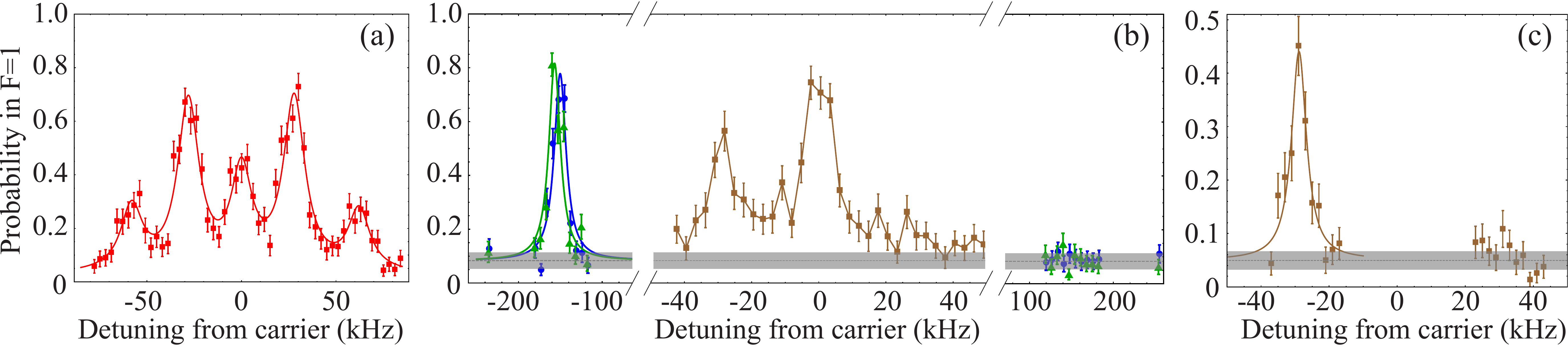}
	\caption{Axial spectra and 3D ground state cooling of a single neutral atom. (a) A thermal axial spectrum (red squares) using an intensity corresponding to a cold carrier Rabi frequency of $12$ kHz and a $\Delta t= 65~\mathrm{\mu s}$ Raman pulse. The data are fit to Lorentzians (solid line) to guide the eye.  (b) Result of simultaneous sideband cooling in three-dimensions, demonstrating significant sideband asymmetries and simultaneous ground state occupations in all dimensions. The axial data (center) illustrates our cooling parameters, and is performed with a carrier Rabi frequency of 10.6 kHz and a pulse of $236~\mathrm{\mu s}$ (near $\pi$ pulse on the ground state $\Delta n=+1$ sideband, a $5\pi$ pulse on the carrier). For the radial data, the blue circles (green triangles) are spectra along the $y'$ ($x'$) axis using a $75~\mathrm{\mu s}$ pulse. The solid lines on the $\Delta n = +1$ sidebands are Lorentzian fits. (c) After 3D cooling, axial spectroscopy for a halved carrier Rabi frequency of $5~\mathrm{kHz}$ and a pulse of $450~\mu s$. Better spectroscopic resolution affirms a large axial ground state occupation.}
	\label{fig:axial}
\end{figure*} 

For the radial dimensions, the PGC allows us to start our Raman cooling reasonably far into the Lamb-Dicke regime with $\eta_{\mathrm{eff}}^{OP}\sim0.3$ ($\bar{n}\sim1.5$ and $\eta^{OP}=0.16$). Like ion experiments, our cooling scheme operates in a pulsed format:  We apply a coherent Raman transition on the $\Delta n = -1$ sideband using an intensity corresponding to a ground state carrier Rabi frequency of $\Omega_c=2\pi \times 26~ \mathrm{kHz}$,  where the $\Delta n = -1$ sideband Rabi frequency is $\Omega_{sb} \sim \eta^R \Omega_c \sqrt{n}$. Therefore, we apply $60~\mathrm{\mu s}$ pulses for 47 cycles and switch to $75~\mathrm{\mu s}$ pulses for the last 3 cycles when the remaining excited state fraction is primarily in the first excited state.  The pulses are applied to each radial axis by alternating Raman beam pairs.  Interspersed between the Raman pulses are optical pumping pulses $90~\mathrm{\mu s}$ long that recycle the atom back to $\ket{2,2}$. Note we do observe cooling of both radial dimensions even if we employ a single pair of cooling beams, indicating there is coupling between the radial dimensions of our anharmonic potential. To assure and verify cooling in both dimensions, however, we cool and probe each axis separately.

After Raman cooling we see a significant asymmetry in the radial sideband spectroscopy due to a large ground state occupation (Fig.~\ref{fig:spectra}(c)). While the $\Delta n = -1$ sideband is suppressed, the  $\Delta n = +1$ sideband has increased in height due to decreased dephasing as the thermal distribution is narrowed. Figure~\ref{fig:spectra}(d) shows Rabi oscillations for both the carrier (black) and $\Delta n = +1$ sideband (blue) transition, the latter of which oscillates slower by a factor of $\eta^R$. Figure~\ref{fig:spectra}(d) further demonstrates the coherence of our motional transitions.   The carrier decays less quickly than in Fig.~\ref{fig:spectra}(b), as expected for colder atoms; the sideband transition decays slightly faster than the cold carrier due to its relative narrowness~\cite{note2}, but note that it maintains high contrast on the first oscillation. To assess our final occupation, we compare the measured signal at the position of the $\Delta n = -1$ transition peak to the measured background level due to any atoms left in $F=1$ and imperfect push-out efficiency.  We find $\{\bar{n}_{x'},\bar{n}_{y'} \} =\{0.05^{+0.05}_{-0.04}, 0.02^{+0.04}_{-0.02}\}$. 

To achieve large three-dimensional ground state occupations, we must cool the weaker axial dimension of the trap where both spectroscopy and cooling are more challenging due to the smaller trap frequency.  Figure~\ref{fig:axial}(a) shows a thermal axial mode spectrum after PGC. The near equality of the $\Delta n = +1$ and $\Delta n = -1$ transitions and the presence of significant second order sidebands (given $\eta^R=0.23$ for this dimension) are consistent with a small initial ground state population.   Assuming an isotropic initial temperature of $12~\mu\mathrm{K}$, we would expect $\bar{n}=8$, which corresponds to a challenging starting point of $\eta^{OP}_{\mathrm{eff}}=1.4$. Further, the smaller trap frequency makes it difficult to spectroscopically separate the carrier and sideband peaks while maintaining Rabi frequencies that are insensitive to technical dephasing.  For the cooling, a large Rabi frequency leads to off-resonant carrier transitions that cause heating, while too small a Rabi frequency leads to smaller transfer efficiencies and slow cooling.

Despite these barriers, we are able to Raman cool in the axial dimension and achieve the significant three-dimensional ground state occupations evidenced in Fig.~\ref{fig:axial}(b) after 33 ms of cooling (see the Appendix for pulse parameters). Unlike the radial directions, we use different parameters for the axial cooling than are ideal for spectroscopy. To highlight this distinction, we show two spectra of cooled atoms. The axial mode spectrum in Fig.~\ref{fig:axial}(b) approximates the parameters used during our cooling, which balance speed with spectroscopic resolution.  While there is a clear sideband asymmetry in the spectrum, the size of the Rabi frequency compared to the axial trap frequency complicates a temperature analysis because off-resonant carrier transitions occur at the frequency position of the sidebands.  This is illustrated in Ref.~\cite{note3} where we show calculated spectra for these pulse parameters, but a full comparison to this calculation is complicated by the partial dephasing we observe.

In Fig.~\ref{fig:axial}(c), we halve the spectroscopy Rabi frequency to sacrifice coherence for spectroscopic resolution. For this Rabi frequency and pulse, the spectrum is dephased and hence we can understand the spectrum simply as a set of multiple Lorentzians.  We extract our axial temperature from the sideband asymmetry in Fig~\ref{fig:axial}(c), assuming the dephasing uniformly affects the two sideband peaks.  For both the radial and axial dimensions we analyze the data attributing all of the signal observed at the $\Delta n = -1$ position to the sideband, thus placing an upper bound on the achieved temperature.  Taking this approach we extract temperatures in all three dimensions, again subtracting off a measured background and assuming a thermal distribution.  The result is $\{\bar{n}_{x'}, \bar{n}_{y'},\bar{n}_{z} \} =\{ 0.02^{+0.07}_{-0.02}, 0.01^{+0.06}_{-0.01} , 0.08^{+0.08}_{-0.06} \} $.

The above occupations indicate we have cooled a single neutral atom to the ground state of an optical tweezer with $97^{+3}_{-11}\%$ probability in the radial plane, and $93^{+5}_{-7}$\% probability in the axial direction, and hence a three-dimensional ground state population of $90^{+8}_{-16}\%$. While this estimate neglects imperfections in our spin preparation and detection, this is not a fundamental impediment to the final temperature. The current limitation to our quoted temperature is the precision of the spectroscopy measurement.  Investigation of tunneling between adjacent tweezer traps would allow qualitatively new ways of analyzing the purity of the ground state preparation, because bosonic enhancement of tunneling rates is contingent upon this preparation.  Importantly, we have measured an upper bound to the heating rate of less than 1 radial vibrational quanta per second, suggesting tunneling experiments are feasible with sufficiently tight, adjacent tweezer traps.  Such experiments would open a new avenue of research for ultracold atoms in optical tweezers, and establish their utility for several directions in quantum information and bottom-up approaches to quantum simulation.

\acknowledgments{We acknowledge T. P. Purdy and M. Feig for helpful input, and G. Downs for technical assistance. This work was supported by The David and Lucile Packard Foundation and the JILA NSF-PFC.  CR acknowledges support from the Clare Boothe Luce Foundation, AMK and BJL from NDSEG and NSF-GRFP fellowships respectively.}

{\em Note added} -- Related concurrent studies independently carried out by another group are described in Ref.~\cite{noteadded}

\section{Appendix}

\noindent\textbf{Optical tweezer trap and single atom loading.}
The optical tweezer is formed by a custom high-numerical objective lens that is designed to be diffraction limited at both $850~\mathrm{nm}$ and $780~\mathrm{nm}$~\cite{note4}. From our measured trap frequencies, we infer a $1/e^2$ beam radius of $\sim 0.76~\mu \mathrm{m}$ and a depth of $1.4~\mathrm{mK}$ for our intensity. Abberations dominated by astigmatism slightly increase our spot size compared to a diffraction limited spot for a 0.6 NA lens. We measure a $1/e$ lifetime of a single atom of about $5~\mathrm{s}$, which can be extended at the expense of our trap load rate. 

We load the tweezer trap from a vapor cell magneto-optical trap (MOT) that has filled for $100-200~\mathrm{ms}$. The MOT consists of two
orthogonal pairs of retro-reflected beams along $x'$ and $y'$, as well as a third pair at approximately $45^\mathrm{o}$ to the $z$-axis and
$10^\mathrm{o}$ above the $y-z$ plane. For polarization gradient cooling (PGC), the same beams are used as for the MOT light and the bias fields are set for zero magnetic field; we zero our fields on the basis of microwave spectroscopy.  The PGC detunings are optimized for each stage of the experiment.  For loading into the tweezer trap, the $F=2-3'$ light is detuned $60~\mathrm{MHz}$ red of the bare optical transition; the PGC to prepare for Raman cooling uses light $23~\mathrm{MHz}$ red of the bare optical transition~\cite{Tuchendler2008}.

All of our data rely on the ability to make spin-sensitive measurements of our $^{87}$Rb atom.  Our procedure is to apply light resonant with the light-shifted $F = 2 \rightarrow 3'$ transition while the atom is in the trap to retain atoms only in $F=1$.  We then image the $F=1$ atoms by applying PGC (including repump) light.  This allows us to collect fluorescence for $25$ to $50~\mathrm{ms}$ while imaging this light onto an CCD camera with $\sim 7\%$ overall efficiency with the same high-NA lens that creates the tweezer trap.  

\vspace{10pt}
\noindent\textbf{3D cooling parameters.}
For the three-dimensional cooling shown in Fig.~\ref{fig:axial}(b),(c) we toggle the Raman laser pulses to address each of the axes of the trap, and between each pulse we insert $90~\mathrm{\mu s}$ of optical pumping. The cooling process in total occurs in 75 cycles:  The first 50 cycles use a Raman pulselength of $\Delta t = 48~\mathrm{\mu s}$ ($ \Delta t = 40~\mathrm{\mu s}$) for the radial (axial) directions, and then $\Delta t = 72.5~\mathrm{\mu s}$ ($\Delta t = 80~\mathrm{\mu s}$) for the final 25 cycles. We use an intensity corresponding to a cold carrier Rabi frequency of 13 kHz for the axial direction, and 31 kHz for the radial dimension.

\clearpage
\begin{widetext}


\section*{Supplementary material}

# SUPPLEMENTARY MATERIAL

**Optical layout.** Figure S1 illustrates the basic optical layout of our experiment. The optical tweezer that provides three-dimensional confinement is created by focusing 852 nm light through a custom built objective lens. It has a working distance of 21 mm, and is corrected for the 0.25 in. fused-silica window of our octagonal vacuum cell. The objective consists of both aspheric and spherical elements. In creating the light for the trap, we incorporate two acoustic-optic modulators (AOMs) that yield deflections in two orthogonal directions. This will allow for future dynamic positioning of multiple traps and individual depth control of these traps in real time. The same objective that creates the traps also collects fluorescence from the trapped atom, which is then imaged on to a CCD array.

**Raman transitions.** The Raman beams are detuned approximately 50 GHz red of the D2 line with an external cavity diode laser, which is passively stable. The intensity of each of the four Raman beams is separately switched with acoustic optic modulators, and RB1 additionally passes through an in-fiber electro-optic modulator (EOM) operating near the hyperfine splitting of 6.8 GHz. RB1, prior to passing through the EOM, is frequency shifted by +10 MHz from RB2, RB3, and RB4. This ensures that only one of the optical sidebands on RB1 is resonant with a Raman process while the other optical sideband is 20 MHz off-resonance, thereby preventing quantum interference effects and transitions as a result of any residual $\sigma^-$ polarization on RB1 [1]. The 10 MHz frequency difference between the two carriers is also sufficiently large so as to avoid parametric heating. As discussed in the text, RB1 is always used in combination with one of Raman beams to address an axis. The beam detunings and polarizations are such that starting in $F = 2$ the atom always stimulated emits into RB1 while absorbing a photon from the other beam in use, which determines the direction of momentum transfer effected with each beam pair. The light shift on the $|1,1\rangle \leftrightarrow |2,2\rangle$ transition is $\sim 90$ kHz for all of the Raman beam pairs, which is dominated by the vector light shift induced by the $\sigma^+$-polarized RB1 beam.

When driving Raman transitions we observe some technical dephasing; we find the dephasing rates increase as the Rabi frequency of the driven transition is reduced. This is seen in the different dephasing rates of the carrier and the $\Delta n = +1$ Rabi oscillations in Fig. 2(d), as well as the dephasing observed in the axial spectra. We suspect decreasing the resonance width increases sensitivity to technical variations in the light-shifted hyperfine resonance. This instability could be attributed to intensity fluctuations in the Raman beams or the optical tweezer vector light shift, which varies across the atomic wavepacket.

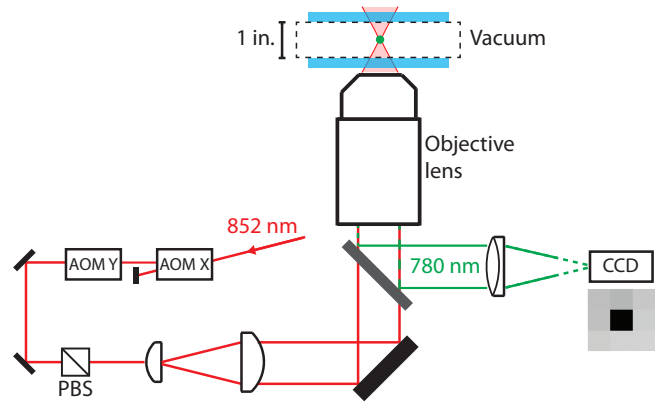

FIG. S1. Optical layout for the optical tweezer trap and atom detection.

**Optical pumping setup and optimization.** The heating during the optical pumping step determines the efficiency of Raman cooling. In our case, this heating rate depends on the recoil per photon and the number of photons scattered to optically pump the atom from $|F, m_F\rangle = |1, 1\rangle$ to $|2, 2\rangle$. Polarization impurity increases the number of photons scattered, and diminishes the fidelity of the pumping.

We take the following steps to optimize the polarization seen by the atoms. The polarization is set with a polarizing beam splitter and a quarter-wave plate. After aligning the optical pumping beam to the atoms, we optimize the direction of the quantization axis to be parallel to the direction of the optical pumping beam. We first optically pump to $|2, 2\rangle$ using $F = 2 - 2'$ light and repumping $F = 1 - 2'$ light. Subsequently, we apply the same $F = 2 - 2'$ light in the absence of the repumping $F = 1 - 2'$ light. If $|2, 2\rangle$ were truly dark, this would have no effect. However, we can detect residual $\pi$ and $\sigma^-$ light via the depumping rate out of $|2, 2\rangle$. Since the $F = 2'$ level decays with equal probability to $F = 1$ and $F = 2$, we can detect depumping via population build up in $F = 1$. As shown in Fig. S2, we then optimize the direction (field angle) and magnitude (tip field) of a small transverse field added to the 3 G quantization axis along the $y$-axis by minimizing this depumping rate.

**Calculation of optical pumping photons scattered.** To calculate the feasibility of the number of cooling cycles required in the experiment, we determine theoretically the number of optical pumping and repumping photons scattered in the ideal situation of pure, resonant $\sigma^+$ $F = 2 - 2'$ light and repumping $F = 1 - 2'$ light [2]. For this calculation, the Hilbert space consists of the magnetic sublevels of $F = 1$, $F = 2$, and $F = 2'$. We perform a master equation calculation, given the atomic line strengths, by numerically solving the differential equa-

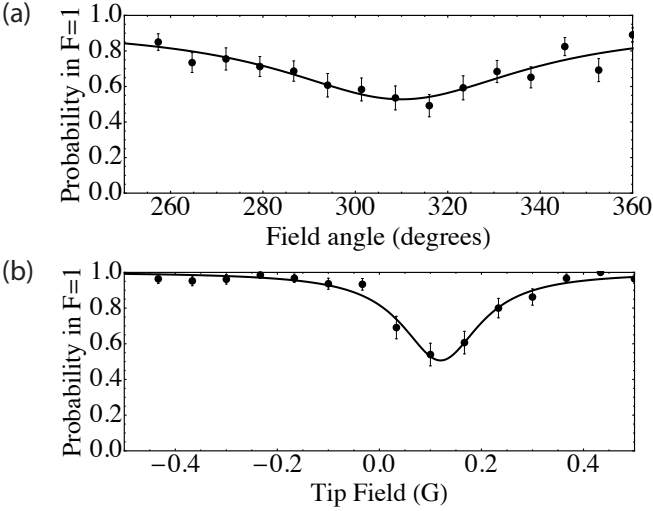

FIG. S2. Optical pumping optimization. After optical pumping to $|2,2\rangle$, we vary the magnetic field quantization axis transverse angle (a) and magnitude (b) and observe the depumped fraction after 4 ms of exposure to $F = 2 - 2'$ light. We attribute residual depumping to imperfect quarter-wave plate alignment, and finite detuning from other transitions on the D2 line.

tion defined by a Lindblad operator $\mathcal{L}$,

$$\mathcal{L} = \sum_{i,j} \gamma_{i,j} \rho \gamma_{i,j}^\dagger - \frac{1}{2}\gamma_{i,j}^\dagger \gamma_{i,j} \rho - \frac{1}{2}\rho \gamma_{i,j}^\dagger \gamma_{i,j}, \quad \text{(S1)}$$

where $\rho$ is a density operator describing the sub-level populations, the $\gamma_{i,j}$ are defined as,

$$\gamma_{i,j} = \Gamma_{i,j}^{1/2}|j\rangle\langle i|, \quad \text{(S2)}$$

and $\Gamma_{i,j}$ describes the decay rate from $|i\rangle$ to $|j\rangle$ due to spontaneous emission, or the incoherent scattering rate of $|i\rangle$ to $|j\rangle$ via the optical pumping and repump beams. From the time evolved density matrix, the number of scattered photons $\langle n_\gamma \rangle$ can be determined from

$$\langle n_\gamma \rangle(t) = \int_0^t dt' \sum_{i,j} \text{Tr}(\gamma_{i,j}^\dagger \gamma_{i,j} \rho(t')), \quad \text{(S3)}$$

where the sum $i$ is taken over the excited state manifold $F = 2'$ and the sum $j$ is taken over the sublevels of the two hyperfine ground states $F = 1$ and $F = 2$. A plot of photon number scattered and population in $|2,2\rangle$ versus time is shown in Fig. S3. We emphasize that $\langle n_\gamma \rangle$ is minimized when the repump light is also $\sigma^+$-polarized.

**Spectroscopy lineshape** For the carrier spectra shown in Fig. 2(a) and Fig. 2(c) we fit to the expected theoretical lineshape. For a two level system driven by a coherent source with Rabi frequency $\Omega$, pulse $\Delta t$, and detuning $\delta$, the transition lineshape is given by the Rabi sinc function,

$$P_{F=1}(\delta) = A\frac{\Omega^2}{\Omega^2 + \delta^2}\sin^2((\Omega^2 + \delta^2)^{1/2}\Delta t/2), \quad \text{(S4)}$$

where $A$ describes an overall amplitude. For the thermal carrier spectra displayed in Fig. 2(a) this model is an approximation because it assumes a pure Rabi frequency. In Fig. 2(c) the Rabi frequency is primarily that of the ground state carrier transition and hence this model is a good description.

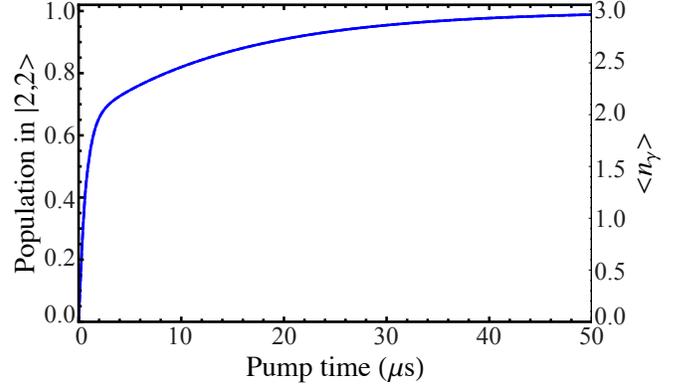

FIG. S3. Optical pumping calculation. Numerical results for optical pumping rate into $|2,2\rangle$ and the number of photons scattered as a function of time. The parameters for this calculation are a repumping rate of 1 MHz, optical pumping rate of 100 kHz, and excited state linewidth of 6 MHz.

**Simulation of coherent dynamics under Raman coupling interaction.** We determine the coherent dynamics in order to better understand our measured spectra and cooling efficiency. We approximate the internal degrees of freedom as a spin-1/2 system, and the external degree of freedom we truncate at 10 vibrational levels. For our initial temperatures in the radial degrees of freedom, the highest level has a population of $\leq 0.005$, and for the axial $\leq 0.05$. Our calculation of spectra and dynamics only considers one external degree of freedom (1D), where the dimension of interest is distinguished by its associated trap frequency. In the rotating frame, the Raman interaction is written,

$$H_\text{R} = \frac{\Omega_0}{2}(\sigma^+ e^{i\Delta k \hat{x}} + \text{h.c.}), \quad \text{(S5)}$$

where $\Omega_0$ is the bare Rabi frequency, $\sigma^+$ is the spin-1/2 raising operator, $\Delta k$ is the momentum transfer for the Raman beam configuration, and the exponential term is the momentum translation operator associated with the coherent momentum kick due to absorption of a photon from one beam and stimulated emission into the other [3]. The argument of the exponential can be recast as follows,

$$i\Delta k \hat{x} = i\eta^R(\hat{a}^\dagger + \hat{a}), \quad \text{(S6)}$$

where,
$$\eta^R = \Delta k x_0, \quad (S7)$$
with,
$$x_0 = \left(\frac{\hbar}{2m\omega}\right)^{1/2}, \quad (S8)$$

the ground state wave-packet size, and $\hat{a}$ ($\hat{a}^\dagger$) the annihilation (creation) operators of vibrational excitations. For numerical calculations, we Taylor expand the momentum-translation operator in powers of $\eta^R$ to 4th order. In the rotating frame, the bare Hamiltonian is,

$$H_0 = -\frac{1}{2}\hbar\delta\sigma_z + \frac{\hbar\omega}{2}|0\rangle\langle 0| + \sum_{n=1}^{9} n\hbar\omega|n\rangle\langle n|. \quad (S9)$$

We solve numerically the combined Hamiltonian $H = H_0 + H_R$, such that for a density matrix $\rho$,

$$\dot{\rho}(t) = -i[H, \rho], \quad (S10)$$

for variable initial conditions. For example, we can solve for the evolution of thermal states,

$$\rho_T = \frac{1}{Z}\sum_{n=0}^{9} e^{-n\hbar\omega/k_b T}|n\rangle\langle n|, \quad (S11)$$

where $Z$ is the partition function associated with a truncation of 10 states.

We can numerically simulate spectra and dynamics with this model. We show in Fig. S4(a) the expected spectrum for an axial thermal distribution characterized by $\bar{n} = 0.1$ and the spectroscopy parameters used in the spectrum in Fig. 3(b); this calculation does not include effects of technical dephasing. For these pulse parameters, we see significant off-resonant carrier contributions for our square pulses, which are comparable to the height of the $\Delta n = -1$ sideband. This behavior supports our discussion in the text, namely that carrier contributions can complicate a temperature analysis.

Now we consider the analysis of the thermal dephasing observed in carrier Rabi oscillations as in Fig. 2(b). While it is possible to use Eqn. S10 to numerically calculate carrier and Rabi evolutions, there are also exact expressions for the Rabi frequencies. We use these expressions to extract a temperature from carrier Rabi evolutions. For a thermal distribution of populated vibrational levels, there is dephasing of the carrier Rabi oscillations because each level has a different associated carrier Rabi frequency to above first order in $\eta^R$. To fit thermal carrier Rabi oscillation data such as that shown in the main text in Fig. 2(b), we use the following model with $\Omega_0$, $T$, and an overall normalization as free parameters,

$$P_{F=1}(t) = \sum_{n=0}^{9} \frac{P_T(n)}{2}(1 - \cos(\Omega(n)t)), \quad (S12)$$

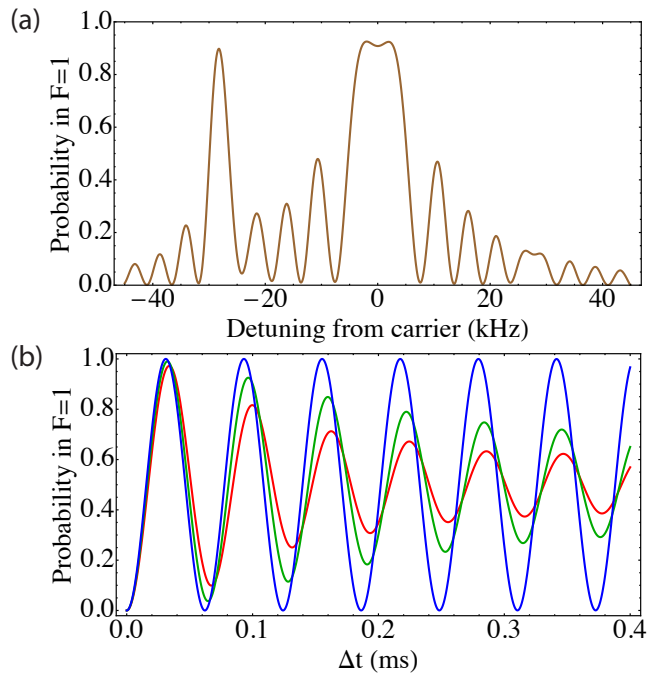

FIG. S4. Coherent dynamics. (a) We simulate an axial spectrum for the parameters: $\bar{n} = 0.1$, $\Omega_0/2\pi = 10.6$ kHz, $\Delta t = 236$ $\mu$s, $\eta^R = 0.23$, and $\omega/2\pi = 30$ kHz. (b) For the radial direction, which exhibits $\eta^R = 0.22$ and $\omega/2\pi = 150$ kHz, we show the carrier Rabi evolution for temperatures of 1 $\mu$K, 10 $\mu$K, and 20 $\mu$K corresponding to the blue, green, and red plots, respectively.

where the level dependent carrier Rabi frequency can be expressed,

$$\Omega(n) = \Omega_0 e^{-(\eta^R)^2/2} L_n((\eta^R)^2), \quad (S13)$$

where $L_n$ is the Laguerre polynomial, and $P_T(n)$ is the normalized Boltzmann factor for a thermal state of temperature $T$. Three examples of Rabi evolutions for different thermal occupations are displayed in Fig. S4(b), which demonstrate increasing dephasing as the thermal distribution broadens [1, 3–5].

**Master equation calculation of Raman cooling.** To understand the expected cooling rate, we model the cooling process with a master equation formalism. According to our optical pumping calculations, we assume there are three photons scattered during the optical pumping step of the Raman cooling. Hence, we define an optical pumping operator as follows. We take the cooling state to be the spin down state, where Raman transitions cause spin-flips to the spin up state. The optical pumping operator for an effective optical pumping rate $\Gamma$ is defined as follows,

$$\hat{O} = \Gamma^{1/2}\sigma^- \left(e^{ik\hat{x}}\right)^3, \quad (S14)$$

where the spin-lowering operator $\sigma^-$ recycles the spin-state while the second term realizes the associated heat-

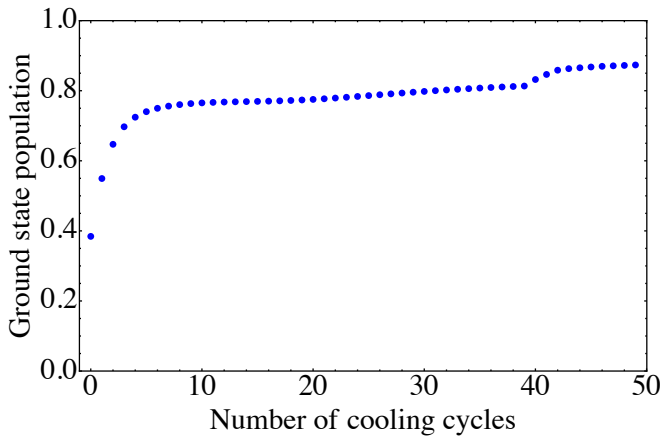

FIG. S5. Cooling calculation. Starting from a temperature of 15 $\mu$K, we simulate the cooling process. We assume here that three photons are scattered during the optical pumping, and that the spontaneous emission process is ignored (only the absorption is accounted for). This corresponds to a 150 kHz trap frequency, as in our radial direction, and a $\Omega_0/(2\pi) = 26$ kHz Rabi frequency. The coherent pulse lengths on the $\Delta n = -1$ sideband are initially 60 $\mu$s, and then 75 $\mu$s for the last 10 cycles.

ing due to three scattered photons of wave-number $k$. We accordingly define the Lindblad operator,

$$\mathcal{L} = \hat{O}\rho\hat{O}^\dagger - \frac{1}{2}\hat{O}^\dagger\hat{O}\rho - \frac{1}{2}\rho\hat{O}^\dagger\hat{O}. \quad (S15)$$

The master equation to be numerically solved is then,

$$\dot{\rho}(t) = \mathcal{L}. \quad (S16)$$

To simulate the cooling, we alternate between evolution under the coherent process defined by Eqn. S10 and the dissipative process defined by Eqn. S16. This accurately models our experiment where the optical pumping and coherent Raman drive are never concurrent, in contrast to a steady state cooling process in which coherent coupling and optical pumping are simultaneous. We can calculate the ground state fraction by finding the expectation value of the dark state projector. An example of a cooling trajectory similar to our experiment is displayed in Fig. S5, from which we conclude that the 50 cooling cycles we use experimentally is reasonable. With better choice of final pulselengths it is possible to increasingly improve the dark state population in simulation, and the disparity in final occupations between data and experiment could be attributed to small differences between the experimental and calculation parameters.


\end{widetext}

\end{document}